\documentclass[twoside]{dis07}
\usepackage[latin1]{inputenc}
\usepackage[dvips]{graphicx,epsfig,color}
\usepackage{wrapfig,rotating}
\usepackage{amssymb,amsmath,array}

\newcommand{\pom}{I\!\!P}
\newcommand{\xpom}{x_{\pom}}

\begin{document}
\title{Status of the H1 Very Forward Proton Spectrometer}

\author{Laurent Favart
%
\thanks{This work is supported by the Fonds National de la Recherche
Scientifique Belge (FNRS).}
%
\vspace{.3cm}\\
%
I.I.H.E., Universit\'e Libre de Bruxelles, CP 230\\
1050 Brussels - Belgium\\
on behalf of the H1 Collaboration
}

\maketitle

\begin{abstract}
 The Very Forward Proton Spectrometer (VFPS) of the H1 experiment at
HERA is collecting data since 2005. The fiber detectors in the Roman
pots located at 218 and 222m downstream from the H1 interaction point,
tag and measure diffractively scattered protons with a high
acceptance in the $\xpom$ range [0.01, 0.025].
 The experimental set up and the spectrometer tagging performance
using diffractive events collected during 2006 and 2007 are discussed.
\end{abstract}

\section{Introduction}

 In recent years considerable progress has been achieved in the partonic
interpretations of
diffractive processes in $e-p$ collisions (see e.g.~\cite{baryons04}),
Most of diffractive studies performed up to now at HERA have been based on
the characteristic presence of a rapidity gap in the diffractive final state.
The precision of this method is limited by the uncertainty related to
the presence of dissociated proton background events. 
The only precise and unambiguous way of studying diffraction is by
tagging the diffracted proton and measuring its four momentum
by means of a proton spectrometer.
Such devices have been used by the H1 and ZEUS Collaborations
and have delivered interesting results,
but their acceptances are small, with the result that the collected
statistics are limited and large systematic errors affect the
measurements.
To fully profit from the HERA luminosity upgrade in the
study of diffraction after the year 2003,
a Very Forward Proton Spectrometer (VFPS) which identifies and
measures the momentum of the diffracted proton with a high
acceptance has therefore been installed by H1.
This contribution reports the VFPS tagging performance
using diffractive events collected during HERA running at high energy
(27.5 GeV for the electron/positron beam and 920 GeV for the proton) 
in 2006 and 2007.
\\

\section{Roman Pot detectors}

 The VFPS \cite{prc} is a set of two ``Roman pots'' located at 118m and 222m 
downstream of the H1 interaction point.
Each pot consists of an insert into the beam pipe,
allowing two tracking detectors equipped with scintillating fibres
to be moved very close to the proton beam.
\\

Many aspects of the design of the Roman pots, including the
stainless plunger vessel and the scintillating fiber detectors,
are adaptations of the FPS proton spectrometer~\cite{fps},
installed and operational in H1 since 1994.
Both detectors of each Roman pot consists of two planes of
scintillating fibres 
oriented at  $\pm 45^0$ w.r.t.\ the horizontal plane and moving
perpendicularly to the beam line direction.
Each detector allows to reconstruct of the position of one
impact point
of the scattered proton trajectory with a precision of about 100 $\mu$m.
 For triggering purposes each detector is sandwiched between 2 scintillating
planes which are connected to different PM's. 
A trigger signal, corresponding to a activity in at least 3 planes
out of four, is delivered separately for each station at the first 
trigger level.
\\

\section{VFPS installation and running}
The VFPS have been installed at the very end of 2003. 
Radiation damage of the optical readout fiber prohibited data taking
during 2004. Hence data available for physics 
\begin{wrapfigure}{r}{0.5\columnwidth}
\centerline{\includegraphics[width=0.45\columnwidth]{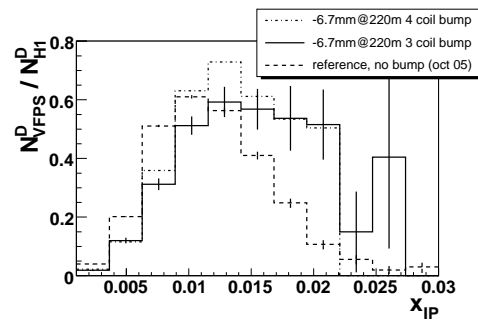}}
\caption{Ratio of events tagged by the VFPS to the  
diffractive events seen in H1 (see section~\ref{sec:dis}) as a function
of $\xpom$ for different beam optics.}
\label{fig:kick}
\end{wrapfigure}
analysis started in 2005. 
The bulk of data were taken in 2006 and 2007, they correspond to and
integrated luminosity of 140 $pb^{-1}$. 
From an operational point of view, the VFPS was into data taking position
for 70\% of the luminosity collected by H1.
\\

The $p$ beam orbit has been modified in April 2006 to increase the VFPS
acceptance. A large fraction of protons with a energy loss above 2\% w.r.t. 
the beam energy are hitting the beam pipe around 200m when the nominal orbit 
is used. The
orbit has been changed moving the $p$ outwards HERA by 6mm at about 200m 
from the interaction point. This procedure leads to an increase of the
rate of events tagged by the VFPS at $\xpom > 0.015$ 
(see Fig.\ \ref{fig:kick}).

\section{Inclusive Diffraction in DIS regime}
\label{sec:dis}
 To study the description of the beam optics and
of the VFPS system by our simulation, first, VFPS tagged events, i.e.
with a fired trigger, are compared to the full sample of diffractive events
selected using the information from the main detector using the rapidity
gap method (see e.g.\ \cite{h1dis07}). 
The full event sample is selected asking for an
electromagnetic cluster in the backward (lepton beam direction) calorimeter 
SpaCal of more than 10 GeV (corresponding to the scattered electron
candidate), a reconstructed vertex and that the most forward particle in the 
main detector has pseudo-rapidity of less than 2.5 (this latest condition 
is equivalent to asking for a rapidity gap). Additionally the Forward
Muon Detector should not have recorded a signal above the noise level.
One can then look to what fraction of this sample largely dominated by
diffractive events is tagged by VFPS. 
This selection was applied to produce the 
Fig.~\ref{fig:kick} discussed in the previous section. 
Over the 140 $pb^{-1}$ collected,
880,000 events are tagged by the VFPS. If a kinematic cut of $Q^2>10$
GeV$^2$ is applied, 215,000 events remain.
\\

\begin{figure}[htbp]
 \begin{picture}(100,120)
  \put(-10,-10){\includegraphics[width=0.52\columnwidth]{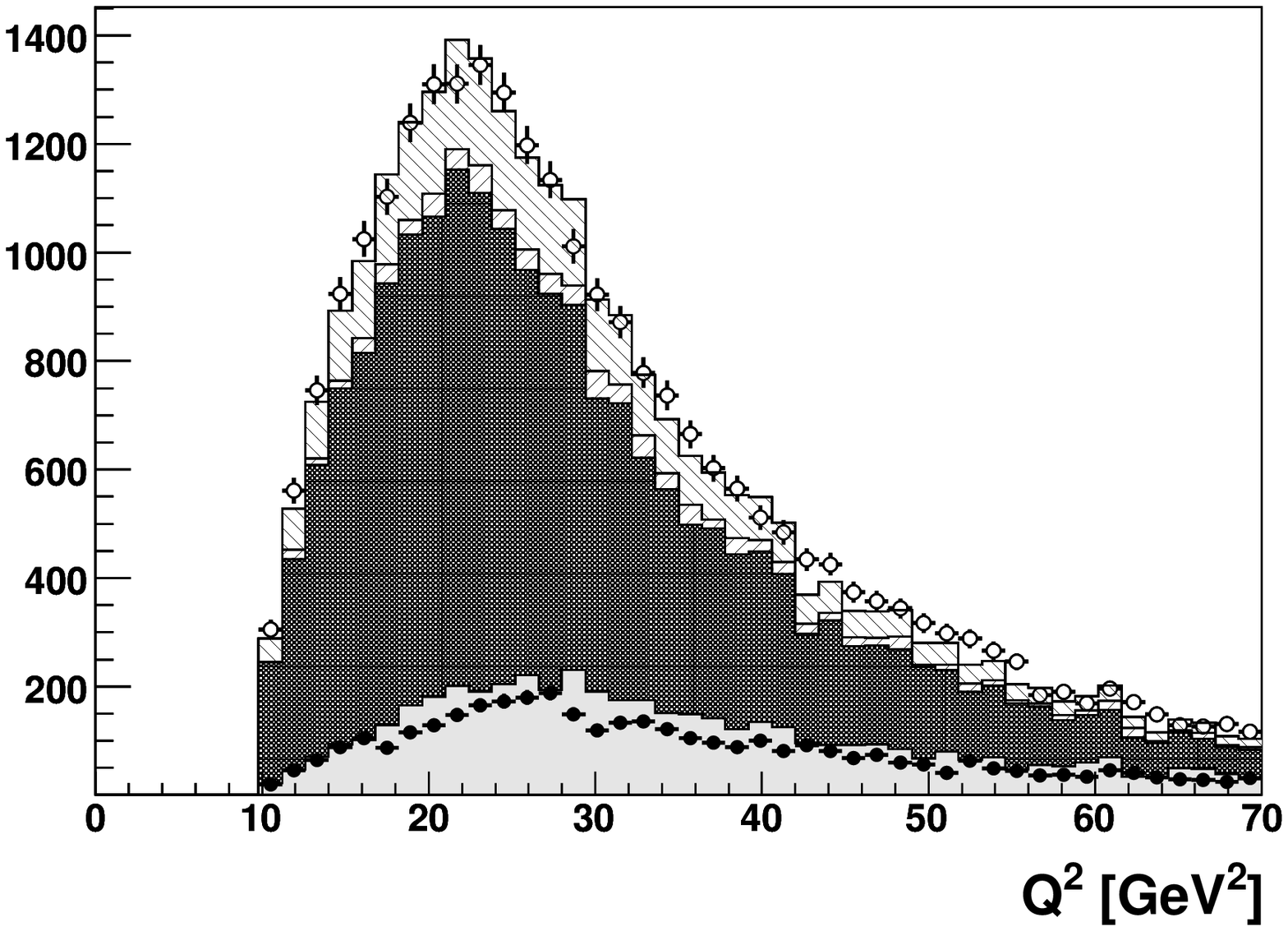}}
  \put(206,-10){\includegraphics[width=0.52\columnwidth]{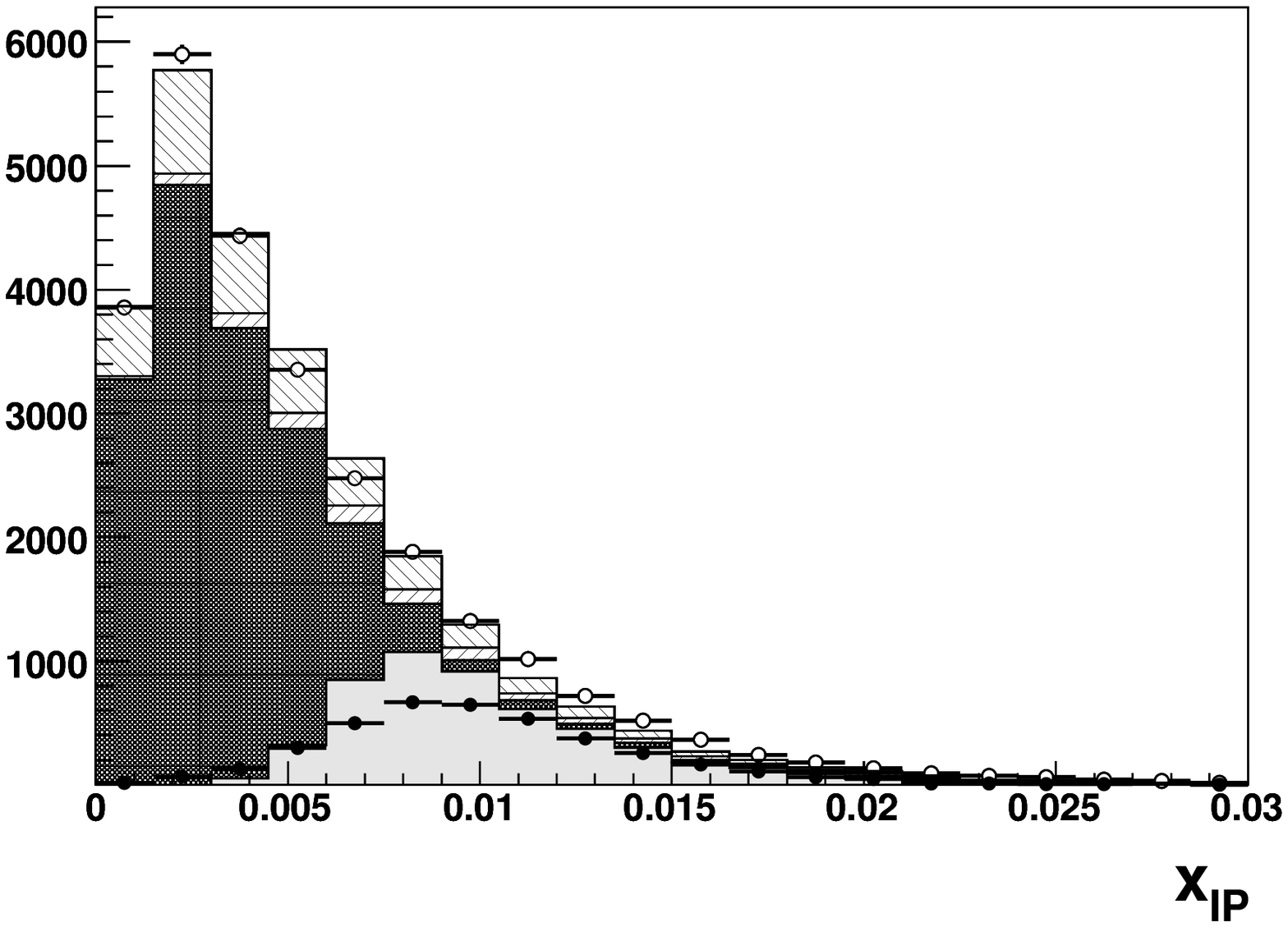}}
  \put(120,69){\includegraphics[width=0.20\columnwidth]{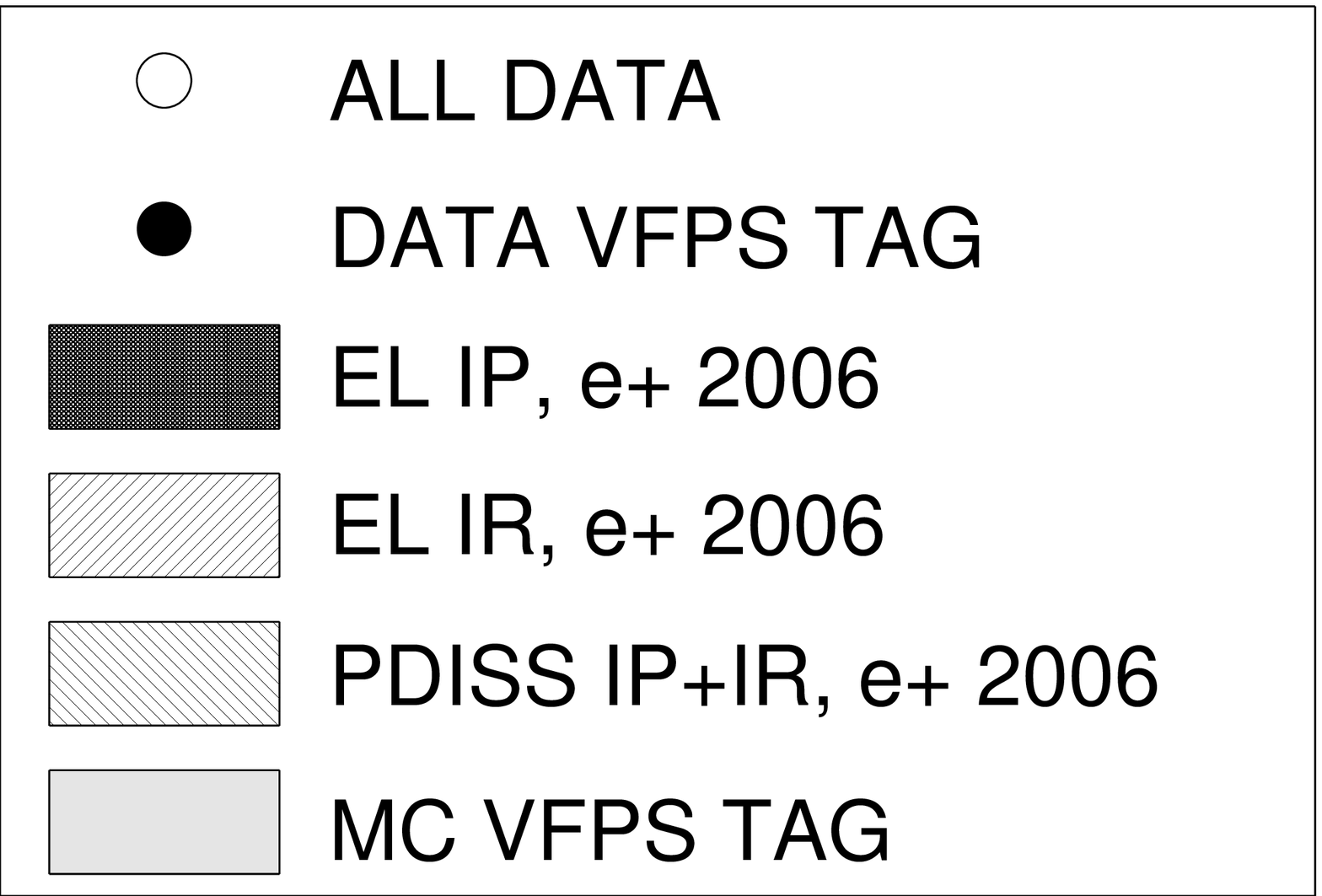}}
 \end{picture}
\caption{Diffractive events in DIS (empty points) are compared to VFPS tagged
events (full points) and to Monte Carlo simulation. {\bf Left:} as a
function of $Q^2$. {\bf Right:} as a function of $\xpom$. The Monte Carlo simulation
contains contributions from Pomeron exchange with elastically scattered proton (EP IP), 
Reggeon exchange with elastically scattered proton (EP IR), 
Pomeron and Reggeon exchange with proton dissociation (PDISS IP+IR). Among them 
simulated events tagged by the VFPS are shown (MC VFPS TAG).}
\label{fig:dis}
\end{figure}

This sample is compared to the sum of diffractive and background
contributions as estimated by Monte Carlo.
In Fig.~\ref{fig:dis} data corresponding to about 1 month of running in 
$e^+p$ mode with a 6mm bump applied (24 $pb^{-1}$)
are compared to Monte Carlo predictions (see figure caption for details). 
\\

 A good agreement is found between the full data sample and the Monte
Carlo (normalized to the data luminosity). 
The trend in $\xpom$ of VFPS tagged sample is described by the Monte
Carlo. A more precise understanding of the VFPS acceptance is
needed and will lead to a better description in $\xpom$.

\section{Diffractive dijets}

\subsection*{Diffractive dijets in DIS} 

 The analysis of diffractive dijets in DIS regime is based on
42.6$e^{-}$ and 54.7$e^{+}$ $pb^{-1}$ taken in 2006.
 Additionally to the selection applied in the previous section, a requirement of at least 
two jets (using the $Kt$ algorithm) is asked, with a minimal transverse momentum in the
photon-proton frame of $p_{T,1}^{*}>5.5 GeV$ and $p_{T,2}^{*}>4 GeV$ respectively for the
first and the second jet. The jets are asked to be well contained in the main detector,
by requirering $\eta_{j1,j2} \in [-1,2]$. \\

\begin{figure}[htbp]
 \begin{picture}(100,80)
  \put(0,0){\includegraphics[width=0.32\columnwidth]{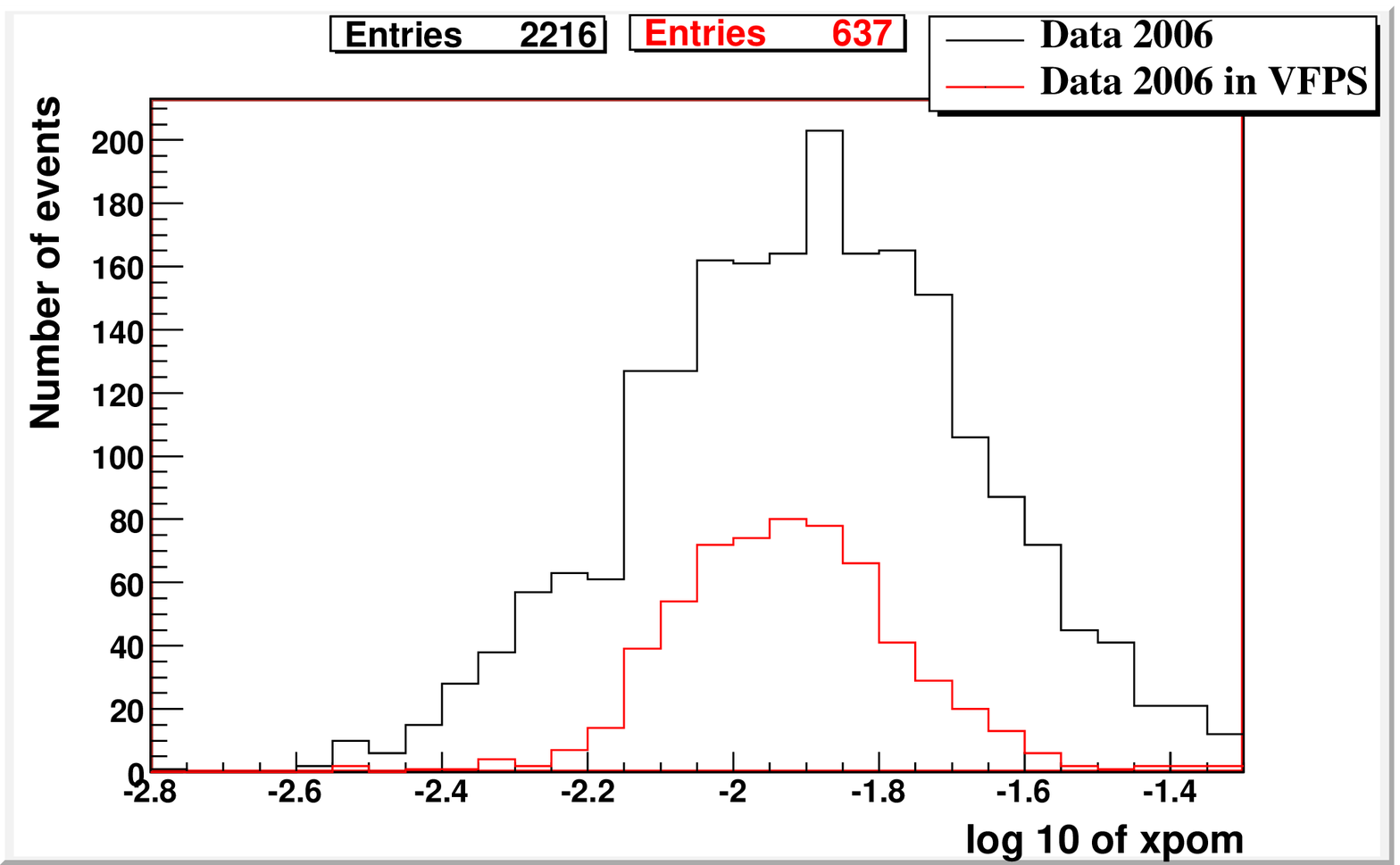}}
  \put(140,0){\includegraphics[width=0.32\columnwidth]{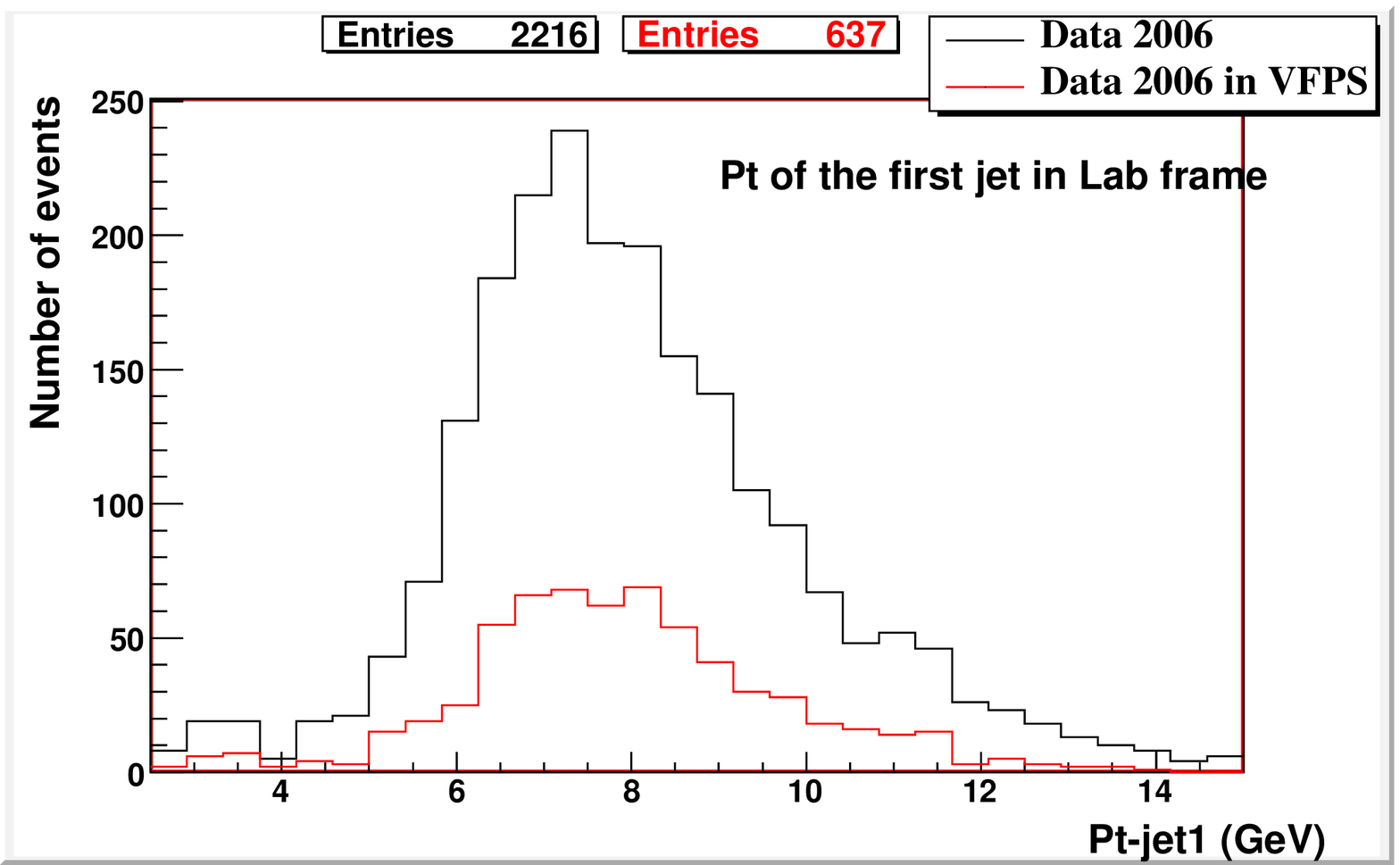}}
  \put(280,0){\includegraphics[width=0.32\columnwidth]{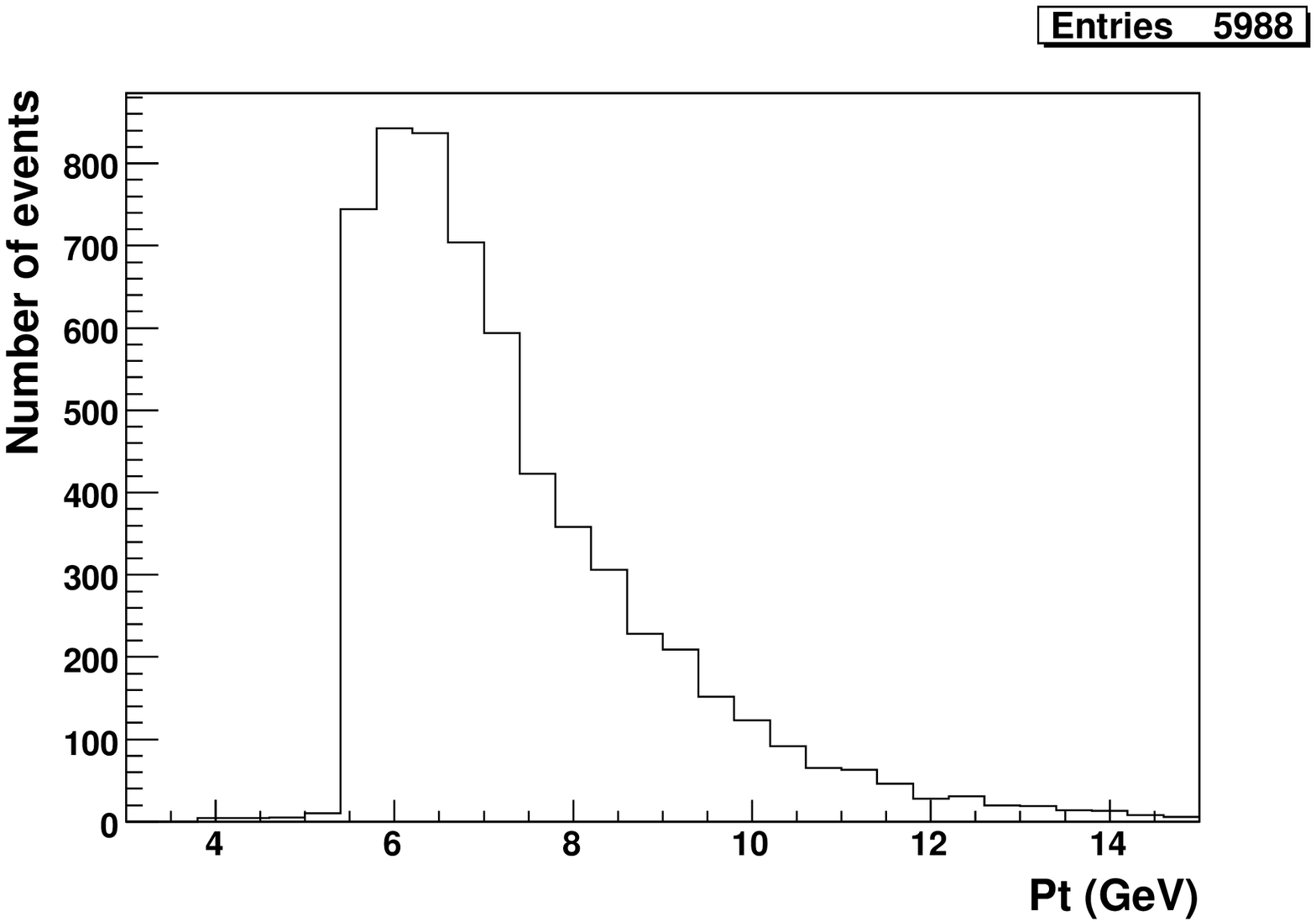}}
  \put(22,58){a)}
  \put(162,58){b)}
  \put(300,65){c)}
 \end{picture}
\caption{{\bf a)} $\xpom$ distribution of dijet diffractive events in DIS.
The highest histogram corresponds to the full dijet sample and the lowest one to the 
VFPS tagged dijet sample. 
{\bf b)} Highest transverse momentum of jets in the laboratory frame of dijet
diffractive events in DIS.
The highest histogram corresponds to the full dijet sample and the lowest one to the 
VFPS tagged dijet sample.
{\bf c)} Highest transverse momentum of jets in the laboratory frame of dijet
VFPS tagged diffractive events in photoproduction.
}
\label{fig:jets}
\end{figure}

The $\xpom$ distribution is shown in Fig.\ \ref{fig:jets}a
comparing the full dijet sample and VFPS tagged dijet sample. This plot illustrates the
well suited acceptance of the VFPS for the dijets production in diffraction.
In Fig.\ \ref{fig:jets}b the transverse momentum of the first jet in the laboratory
frame is shown. Here again the full dijet sample is compared to the VFPS tagged dijet
sample. 

\subsection*{Diffractive dijets in photoproduction} 
To record with a high efficiency diffractive dijet events in photoproduction tagged by
the VFPS, a special trigger has been developed. It allows to lower the threshold 
in jet transverse momentum down to 5 GeV.
A luminosity of 23.7 $pb^{-1}$ has been collected in 2006 and 2007 with that trigger 
corresponding to a selected sample of 6000 events. 
The selection criteria are the same as in the dijet DIS case except that the
scattered electron escapes undetected, at small angle, in the beam pipe.
Figure \ref{fig:jets}c shows the transverse momentum of the first jet in the laboratory
frame (equivalent to the photon-proton frame for the transverse
direction in the present photoproduction case). The distribution of VFPS tagged events
cannot be compared to a full dijet sample as no trigger allowed to keep efficiently
those events down to a transverse momentum of 5 GeV.

\section{Conclusion}

 The VFPS has run successfully collecting a luminosity of 140 $pb^{-1}$. The observed
acceptance is high (above 60\%) in a region of $\xpom$ around $10^{-2}$. The trend in
$\xpom$ of diffractive events tagged by the VFPS is described by the Monte Carlo simulation. 
Important statistics have been collected (880,000 diffractive DIS events, 800 dijets
diffractive DIS events and 6000 dijets diffractive events in photoproduction) for
diffractive structure function measurement and QCD factorisation tests. The proton momentum
reconstruction based on VFPS fiber information is still in progress.  






\begin{footnotesize}




\begin{thebibliography}{99}
\bibitem{baryons04} L.\ Favart, {\it Experimental review of diffractive phenomena},
in proceedings of the 10$^{\rm th}$ International Baryons Conference (BARYONS 2004), 
[hep-ex/0501052].
\bibitem{prc} {\it Proposal for Installation of a Very
Forward Proton Spectrometer in H1 after 2000}, Proposal submitted to the
Physics Research Committee, {\bf PRC-01/00}, H1 note {\bf H1-05/00-582},
{\sf http://www-h1.desy.de/h1det/tracker/vfps/}

\bibitem{fps}
{\it Upgrade of the H1 Forward Spectrometer},
{\bf PRC-96/01}

\bibitem{h1dis07}
J.\ Lukasik, these proceedings.

\end{thebibliography}
%

\end{footnotesize}


\end{document}